\documentclass[twocolumn,showpacs,amsmath,amssymb]{revtex4}
\usepackage{graphicx}
\usepackage{dcolumn}
\usepackage{bm}
\usepackage{braket}
\usepackage{amsmath}
\usepackage{amssymb}
\usepackage[dvips]{color}

\begin{document}
\title{Modeling the nanoscale linear response of superconducting thin films measured by a scanning probe microwave microscope}

\author{Tamin Tai$^{1,2}$}
\author{B. G. Ghamsari$^{2}$}
\author{Steven M. Anlage$^{1,2}$}

\affiliation{$^{1}$Department of Electrical and Computer
Engineering, University of Maryland, College Park, Maryland 20742-3285, USA}

\affiliation{$^{2}$Department of Physics, Center for Nanophysics
and Advanced Materials, University of Maryland, College Park, Maryland  20742-4111, USA}

\date{\today}

\begin{abstract}
A localized and strong RF magnetic field, created by a magnetic write head, is used to examine the linear electrodynamic properties of a Nb superconducting film. The complex reflection coefficient of the write head held in close proximity to the films is measured as a function of sample temperature. A model combining a magnetic circuit (magnetic write head inductively coupled to the sample) and transmission line (microwave circuit) is given to interpret the linear response measurement. Additionally, this reflection linear response measurement can be used to determine the temperature dependence of the magnetic penetration depth on a variety of superconductors.
\end{abstract}

\maketitle

\section{Introduction}

Investigation of superconducting properties such as magnetic penetration depth ($\lambda$), critical temperature ($T_c$) and critical field ($H_c$) is very important because it yields valuable information about superconducting materials for a variety of applications. Generally, the critical temperature and critical field of a superconductor can be measured straightforwardly by performing a standard current-voltage (I-V) and magnetization-field (M-H) measurement, respectively. For the magnetic penetration depth $\lambda$ of a superconductor, various experimental methods have been developed and are summarized in Ref \cite{Talanov} in detail. For example, in order to get an accurate estimation of the absolute value of $\lambda$, Josephson tunneling junctions can be fabricated for a measurement of the Fraunhofer diffraction pattern \cite{TinkhamBook}\cite{Barone}\cite{Cunnane2013}. The main disadvantage of this method is the requirement of fabrication of tunnel junctions for each material of interest. Another precision method is to measure the resonant frequency of a superconducting resonator structure such as the end-plate resonator \cite{Klein1989}, parallel plate resonator \cite{Talanov} or dielectric resonator \cite{Klein1992}\cite{Wilker1992}\cite{Kobayashi} and then calculate the change in absolute $\lambda$. One limitation of this method is that it requires a model for the temperature dependence of the penetration depth, or an independent determination of the absolute value of $\lambda$ at some temperature. For both of these methods, the absolute $\lambda$ is assumed to be a global value common to the entire sample. This assumption is often faulty, especially for newly-developed materials with inhomogeneous properties, or for well-established materials containing defects. We seek to go beyond these simple characterization techniques and develop a spatially-resolved measurement of electrodynamic properties.

Our objective is to examine thin films for use in future superconducting radio frequency (SRF) accelerator cavities, and to understand which defects limit their performance at high RF fields. A reflection method combined with scanning probe technology is used to determine the local $\lambda$ of superconducting materials and to study the homogeneity of the electrodynamic properties. For example, the technology of magnetic force microscopy has been used to study the local $\lambda$ in the pnictide and unconventional superconductors recently \cite{Moler2010}. Additionally, scanning SQUID susceptometry has been applied to precisely measure the change of local $\lambda$ for novel unconventional superconductors \cite{Moler2009}\cite{Moler2012}. However none of the techniques described above can be applied to SRF materials at high RF field and in the GHz regime \cite{Padamsee}. The field generated by SQUID susceptometry is not strong enough and spatial resolution degrades at frequencies in the GHz range \cite{Wellstood}, limiting its application in high field and high frequency study. It is of interest to study SRF materials under conditions of strong RF field excitation, approaching the critical field. Our objective here is to examine the homogeneity of the fundamental electromagnetic properties of candidate SRF materials such as $T_c$ and $\lambda$ under localized excitation at the operating conditions of SRF cavities to understand the drop of quality factor (Q) of SRF cavities in the medium-field and high-field regimes \cite{Padamsee}\cite{Ciovati2010}. The promising candidate materials for SRF cavities are Nb films and MgB$_2$ films \cite{Xi2009}\cite{Tajima} in addition to bulk Nb \cite{Padamsee}.

Based on this motivation, we developed a
novel scanning probe microscope by combining
a magnetic writer and a near field-microwave microscope \cite{Tai2011}\cite{Tai2012}. The magnetic writer
can create a localized and strong RF magnetic field on the surface of the superconducting samples and should achieve surface fields $B_{surface} \sim 10^2$ $mT$ and sub-micron resolution at a frequency of several GHz \cite{Tai2013}\cite{Tai2014}. The design concept is similar in spirit to previous studies of magnetic write heads used to measure ferromagnetic materials \cite{Schultz1996}\cite{Yamamoto1996}\cite{Yamamoto1997}.  What we measure is a reflection signal from the localized area of the superconductor by performing an $S_{11}$ measurement, which is the ratio of the reflected voltage to the incident voltage, with a vector network analyzer (VNA). Due to the temperature (T) dependent magnetic penetration depth $\lambda(T)$ and surface resistance of the measured materials, the reflected voltage will also change with temperature. Although a similar measurement was done on the high $T_c$ cuprates by loop probes with a loop diameter on the millimeter scale \cite{Pond}, our measurement uses a magnetic write head on the sub-micron scale which will generate more intense and localized magnetic fields on the surface of the superconductor. Therefore, many field-dependent parameters, such as the surface reactance, can be measured by applying different RF magnetic fields at different temperatures up to a scale on the order of the thermodynamic critical field $H_c$. The change of $S_{11}$ under different RF magnetic fields and temperatures allows one to calculate the change in surface reactance as a function of the RF field and temperature. In this way, the temperature and RF field-dependent penetration depth $\lambda$ also can be obtained.

\section{Experiment}
\subsection{Linear Response Setup}
\begin{figure}
\centering
\includegraphics [width=3.0 in]{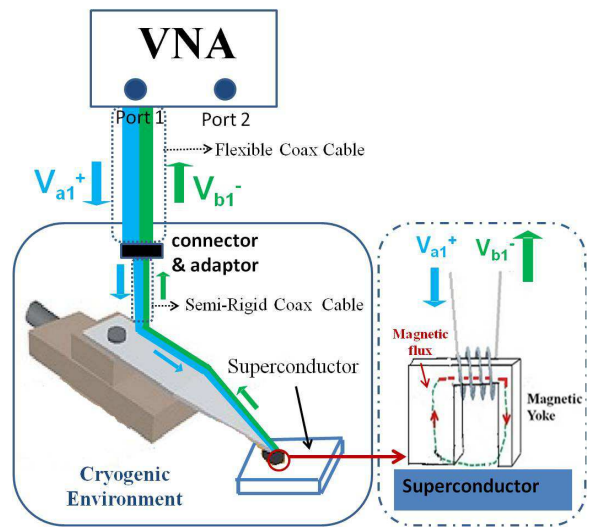}
\caption{Schematic diagram of the linear response measurement, $S_{11}$,
 performed as a function of sample temperature with the vector network analyzer (VNA). The reflected signal $(V_{b1}^{-})$ and incident signal $(V_{a1}^{+})$ are at the same frequency. The inset shows a schematic model of the magnetic write head.}
\label{S11Setup}
\end{figure}

The schematic setup for the linear response measurements is shown in Fig.\ref{S11Setup}. A microwave fundamental tone ($V_{a1}^{+}$) at a given frequency is sent from port 1 of the vector network analyzer (VNA) (model $\sharp$ Agilent
N5242A) into the magnetic writer (from Seagate PINNACLE heads manufactured in 2010 for perpendicular recording technology). The linearly reflected signal (at the same frequency), $V_{b1}^{-}$, in port 1 of the VNA is measured and the complex reflection coefficient $S_{11}=V_{b1}^{-}/V_{a1}^{+}$ is determined.

The process is now discussed in more detail. The magnetic writer is used as a scanning probe
which is integrated into our microwave circuit by soldering the probe
assembly on to a coaxial cable. For this probe, the main part of the writer is a magnetic yoke
surrounded by a several-turn helical coil which transforms the incident signal into
magnetic flux. The yoke is made of a high permeability material (usually ferrite) to
channel the magnetic flux to the narrow gap. It is also shielded to
define a nano-scale bit in the recording medium during the writing process \cite{ReadRiteCorp}\cite{Koblischka2007}\cite{Koblischka2010}. A close-up schematic view of the magnetic write head probe on a superconducting sample is shown on the side of Fig. \ref{S11Setup}. In our design, the
magnetic writer approaches the surface of the superconductor in the range of 200 $nm$ $\sim$ 2 $\mu m$. The fundamental tone stimulates the magnetic writer to generate an RF magnetic field and therefore
excites a screening current on the sample surface so that it can
maintain the Meissner state in the bulk of the material. Larger
magnetic field induces stronger screening current within the
penetration depth ($\lambda$) of the superconducting surface, until
the field reaches the critical field of the material. The time
dependent screening current on the superconductor will induce an
electromotive force (emf) back on the magnetic writer at the same frequency as the excitation. (Note that we do not measure a signal from the magnetic reader element of the write head because it is too far away from the localized excitation on the surface of the superconductor.) The emf
will couple with the incident fundamental tone and reflect back as
an output signal ($V_{b1}^{-}$). We measure the ratio of
$V_{b1}^{-}$ to $V_{a1}^{+}$ ($S_{11}$) at different
temperatures of the superconducting samples. The sample temperature
is controlled by a Lakeshore 340 temperature controller.

Many single-position measurements on different superconducting
samples were performed.
Before cooling down to measure the superconducting reflection signal, an electronic calibration (Ecal) is performed at the end of the flexible coaxial cable at ambient temperature and ambient pressure (black line labeled ``connector $\&$ adapter'' in Fig. \ref{S11Setup}). A part of semi-rigid coax cable on the probe assembly cannot be calibrated with Ecal in the measurement. After drawing a vacuum, the cooling procedure is followed and it takes several hours to stabilize temperature down to almost 4.2 K (without pumping on the helium exhaust), and then the measurement is performed.

\subsection{superconducting samples}
The superconducting samples we study are Nb thin films with
thickness 50 nm. Nb films are made by sputtering Nb onto 3 inch diameter quartz
wafers at ambient temperature under $1.5*10^{-3}$ torr argon pressure. After the deposition, the wafer is diced into many 10*10
$mm^2$ pieces but otherwise left un-disturbed. The $T_c$ of each piece is 8.3 K, which is measured by performing a standard DC resistance measurement. The tested sample is well
anchored to the cold plate to ensure that the surface temperature of
the Nb superconductor is close to that of the cold plate.
The probe is held by a three axis translatable arm. Hence
different points on the surface of the sample can be examined in the same cool-down. We
tested multiple locations on 3 Nb films, and all pieces show consistent
results for their linear response measurement.

\section{Linear Response Measurement}

\subsection{Frequency Response of the Magnetic Probe Assembly}

\begin{figure}
\centering
\includegraphics [width=3.3 in]{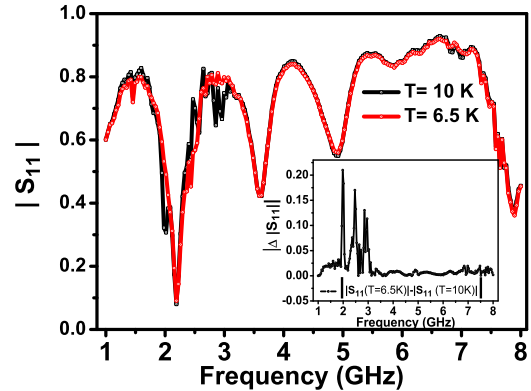}
\caption{Frequency dependence of $|S_{11}|$ in linear scale measured at two different temperatures by the PINNACLE probe. The sample under test is a Nb thin film with thickness 50 nm and $T_c$=8.3 K. The excitation power in both measurements is -15 dBm. The inset shows the absolute value of difference between $|S_{11}|$ at 6.5 K and $|S_{11}|$ at 10 K.}
\label{S11Freq}
\end{figure}

In the high frequency regime, impedance matching of the magnetic probe to the system impedance is a key issue to maximize signal level at the sample. Many different probes were tested and show different suitable operating frequency regimes to satisfy the matching condition for the microwave microscope system \cite{Tai_thesis}. Additionally, different probes will have different resonant frequencies after integrating the probe to the microwave system. The resonant frequency of the probe is also a height dependent parameter due to the variable degree of coupling between the sample and the probe at different separations.
Fig. \ref{S11Freq} shows the frequency dependent $S_{11}$ measurement of the PINNACLE probe under -15 dBm (considered low power) excitation and at a single-position on one of the Nb films. The measurements are taken at two different temperatures: one is in the Nb normal state (10 K) and the other is in the Nb superconducting state (below the $T_c$ of the film). The probe-sample separation, which can be approximately judged by the resonant frequency of the probe assembly, is estimated to be on the order of $0.2 \sim 1$ $\mu m$ from our previous measurements and modeling on many superconducting thin films \cite{Tai2012}\cite{Tai2013}. First, one can clearly see that the deepest dip in $|S_{11}|$ vs frequency happens around 2 GHz, at which the maximum rf magnetic field at the sample can be generated. Second, the $|S_{11}|$ amplitude of the deepest dip is very sensitive to the superconducting transition temperature of the Nb films. From the inset of Fig. \ref{S11Freq}, one can see a significant change of $S_{11}$ amplitude $(\bigl| \Delta \left| S_{11}\right| \bigr|)$ around the dip frequency when the sample becomes superconducting. This measurement can determine the best excitation frequency and the maximum sensitivity at a specific height while the probe approaches the superconducting sample.

\subsection{Temperature Dependence of $S_{11}$ on Nb Thin Film}

\begin{figure}
\begin{quote}
\center
\includegraphics*[width=3.0in]{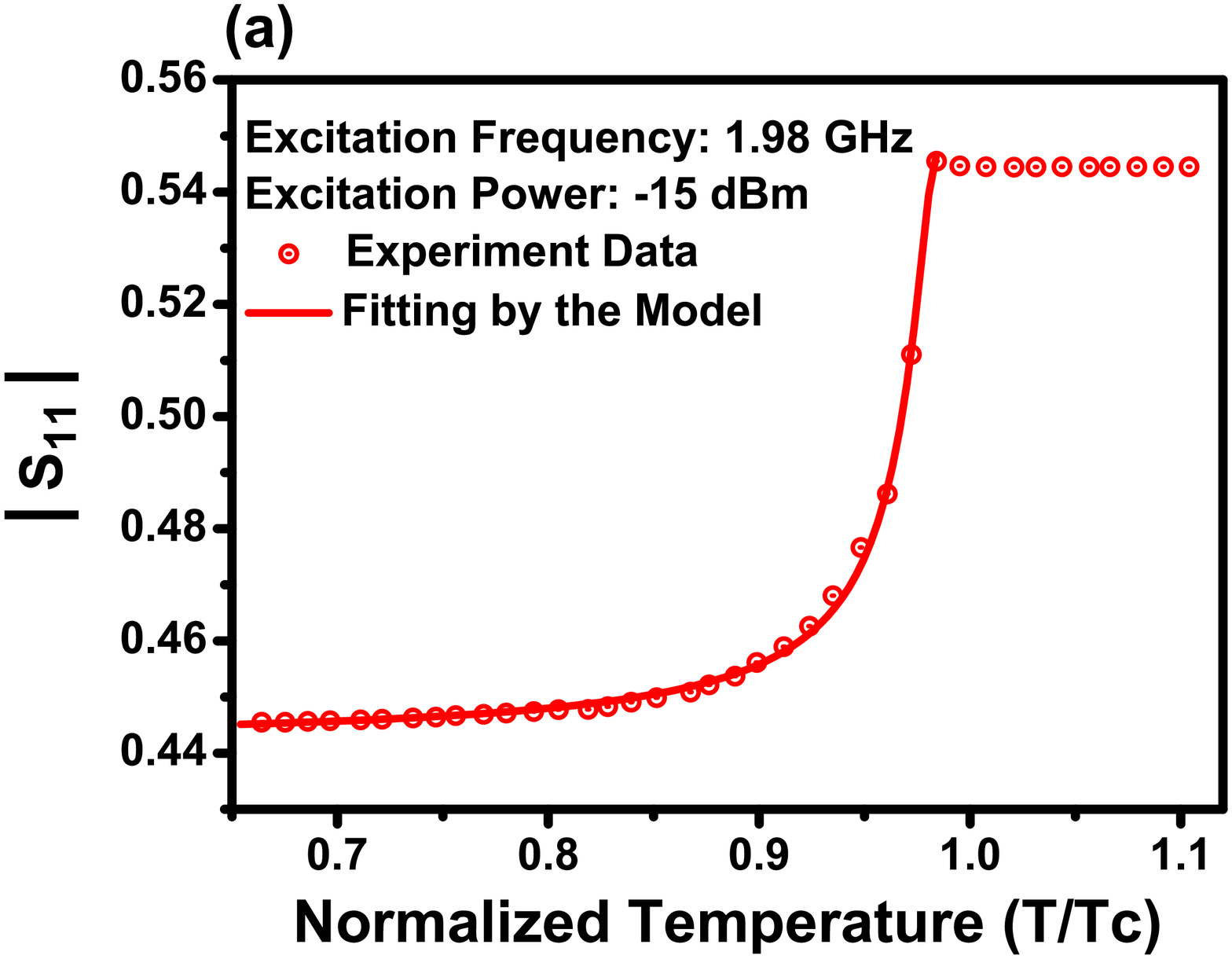}
\includegraphics*[width=3.0in]{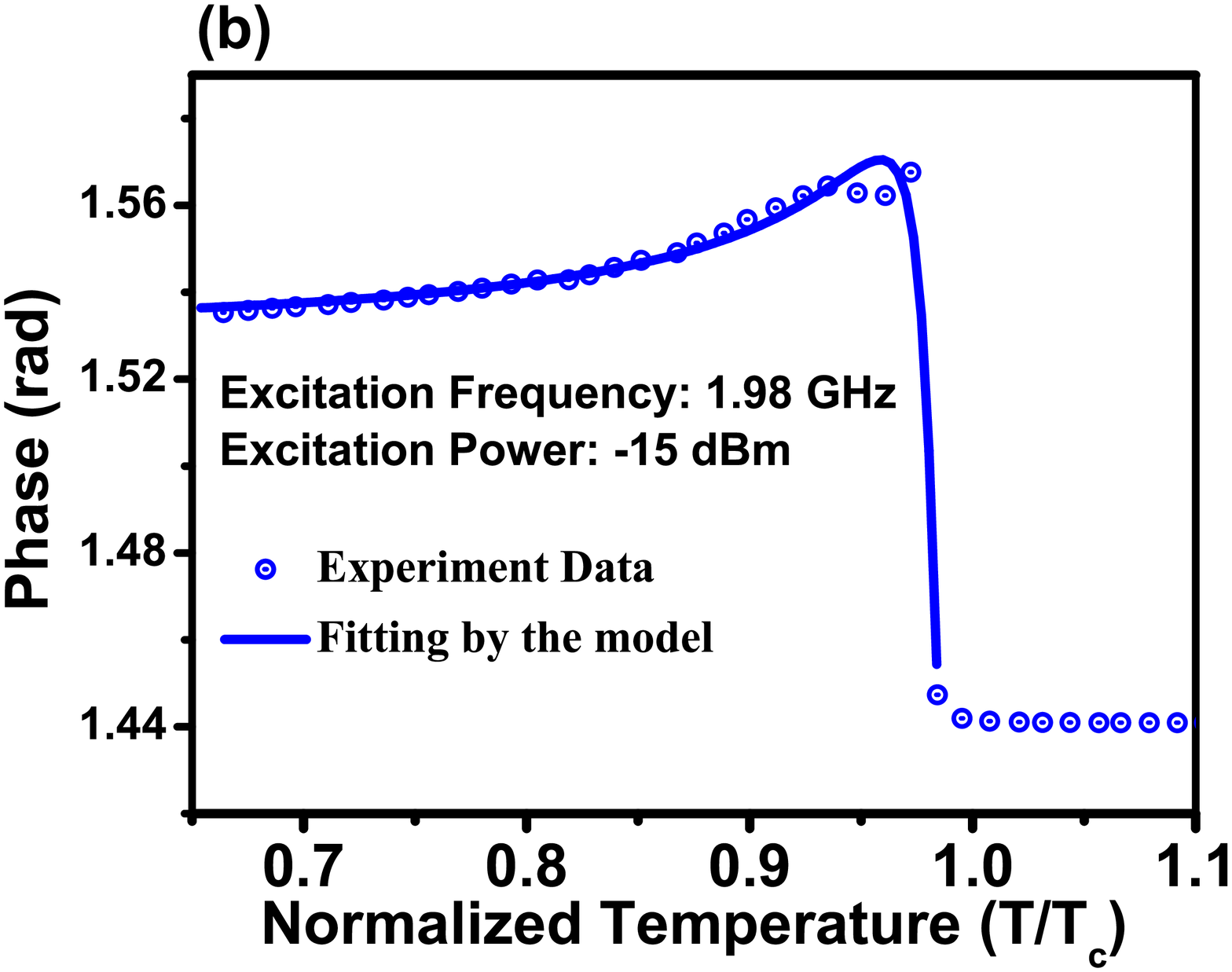}
\caption{(a)Experimental amplitude of temperature dependent $S_{11}$ of a 50 nm Nb thin film is plotted in dot circle symbols measured with the PINNACLE probe at 1.98 GHz, -15 dBm excitation. Note $|S_{11}|$ is plotted on a linear scale. The solid line indicates the theoretical amplitude of $S_{11}$ based on the model described in the text. (b) Dot circle symbols are the corresponding temperature dependent phase of the experimental $S_{11}$. The solid line indicates the theoretical phase of $S_{11}$ based on the model.  Both theoretical curves in (a) and (b) are plotted with $\alpha l$=0.245 (unit-less), $Z_1$=30.23 $\Omega$, $R_{Bkd}$=320.48 $\Omega$, $X_{Bkd}$=-105.13 $\Omega$, $N/R_g$=$6.59*10^{-9}$ $ Henry$ and $\mu_0 /\mu_f$= -0.46 under constant values $nl$=0.065 $m$, $d_{yoke}$=500 nm and $T_c$=8.3 K. }
\label{NbS11T}
\end{quote}
\end{figure}

The dot-circle points in Fig. \ref{NbS11T} (a) and Fig. \ref{NbS11T} (b) shows the amplitude and
phase of a temperature dependent $S_{11}$ measurement, respectively, at a
single-position on one of the Nb films under -15 dBm, 1.98 GHz
excitation by the PINNACLE probe. Note that the excitation frequency in the measurement is decided by the resonant dip in the frequency dependent $S_{11}$ measurement as shown in Fig. \ref{S11Freq}. One can see a sharp change of $S_{11}$ that occurs around normalized temperature $T/T_c=1$ for both the amplitude and phase. This change indicates the Nb transition
temperature ($T_c$). Physically, at this temperature, the Nb thin
film becomes superconducting and generates a strong screening
current on the sample surface which couples with the incident
voltage ($V_{a1}^{+}$) and results in a change of the reflected
voltage ($V_{b1}^{-}$). In other words, from the normal state to the
superconducting state, the surface reactance suddenly changes at the
superconducting critical temperature. The solid curve in Fig.\ref{NbS11T} (a) and Fig. \ref{NbS11T} (b) are the outcome of complex $S_{11}$ curve fitting for amplitude and phase respectively based on the model that will be provided below to quantitatively interpret the temperature dependent $S_{11}$ measurement.

\section{Modeling the Linear Response Measurement}

\begin{figure}
\center
\includegraphics*[width=1.8in]{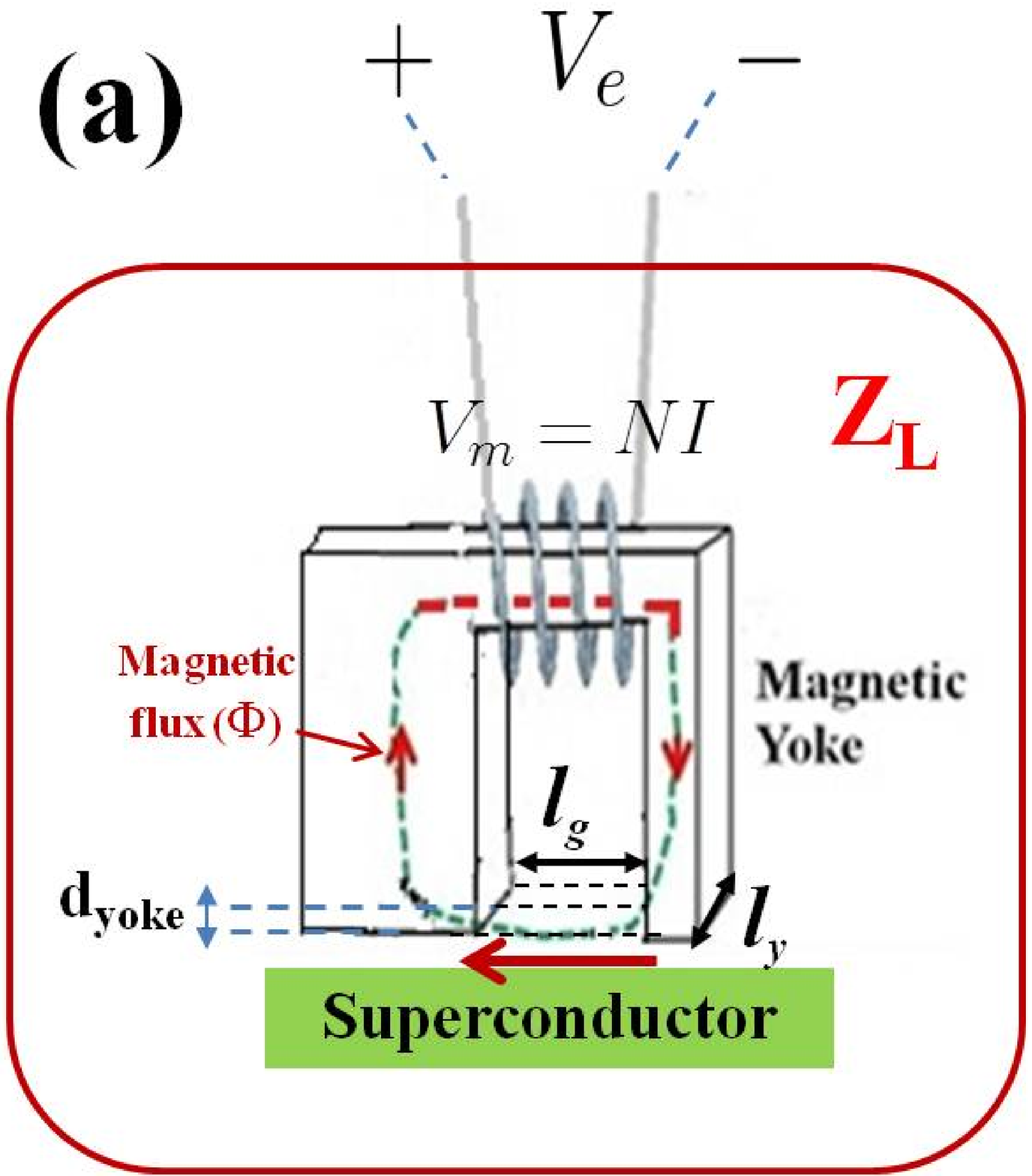}
\includegraphics*[width=1.3in]{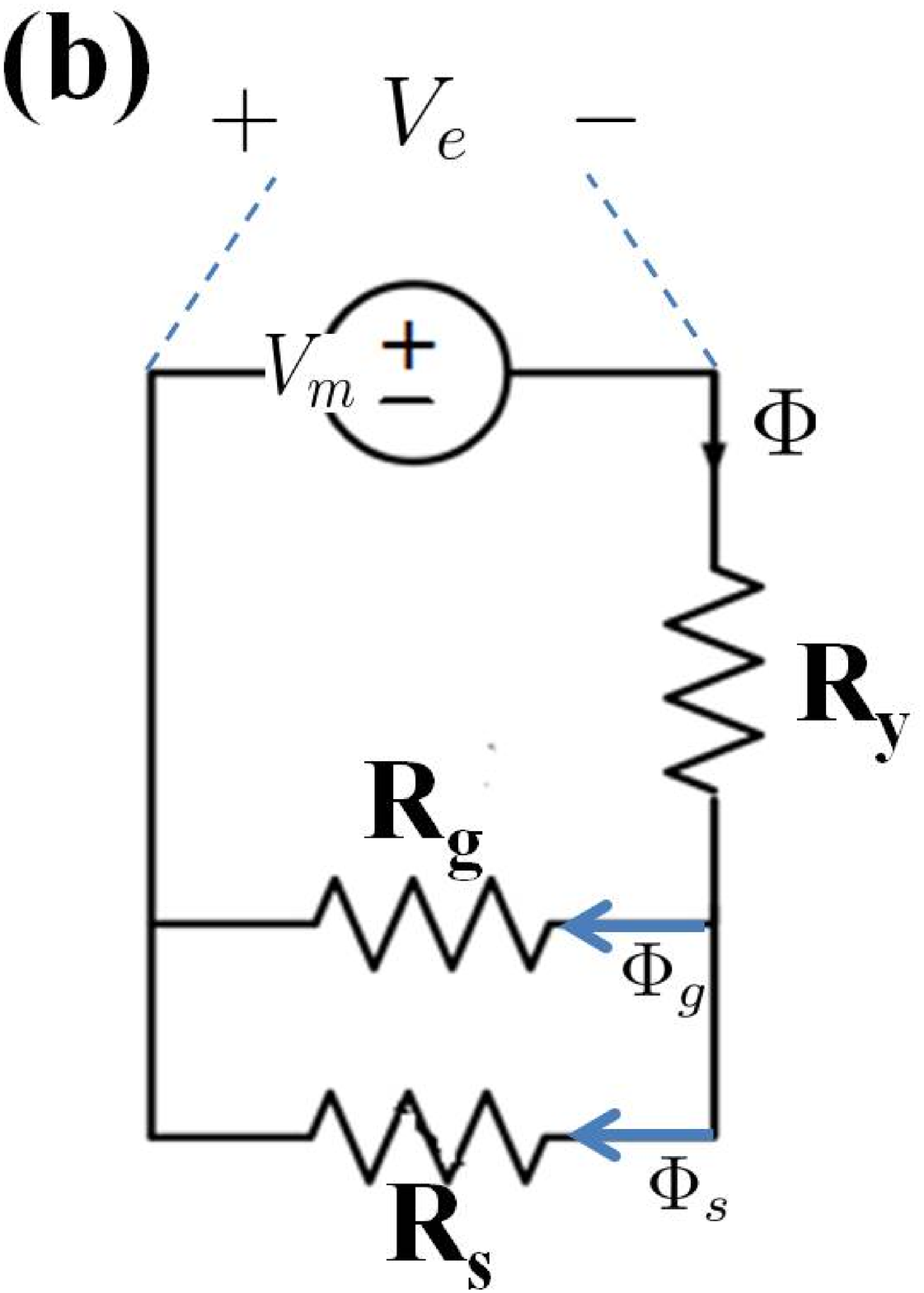}
\caption{(a) A schematic of the experiment where a magnetic probe locally excites the surface of
a superconductor with a time-varying magnetic field. The probe can be modeled as an electrical load impedance, $Z_L$. This figure uses the ring shaped magnetic yoke to schematically represent the writer. $l_g$ indicates the length of the gap under the bottom of the magnetic yoke. $d_{yoke}$ is the effective thickness of the stray magnetic flux in the air gap and $l_y$ is the size of the yoke in the transverse dimensions. $V_m$ is the impressed voltage created by the N turn solenoid with input current I. $V_{e}$ is the induced electric voltage on the input winding.
(b) The equivalent $\textbf{magnetic circuit}$ of the experiment with a superconducting sample and the magnetic probe. $\Phi_g$ is the magnetic flux going through the gap reluctance $R_g$ and $\Phi_s$ is the magnetic flux going through the superconducting sample with reluctance $R_s$ in the Meissner state. $\Phi$ is the total flux in the magnetic circuit. $V_m$ and $V_{e}$ have the same description as in Fig. \ref{ZL_Model} (a).}
\label{ZL_Model}
\end{figure}

For modeling this $S_{11}(T)$ measurement, a Meissner state in the superconductor is assumed while the temperature is below $T_c$. The magnetic yoke is excited by a current in the N-turn solenoid to channel a magnetic flux, $\Phi$, down to the gap as illustrated in Fig. \ref{ZL_Model} (a). The separation between the probe and the sample is assumed to be small compared to the magnetic gap length $(l_g)$, and is ignored in the model. While exciting the surface of a superconductor by the magnetic probe, there are two channels for flux in the magnetic circuit: the air gap and the penetration depth of the superconductor. Since the total flux, $\Phi$, provided by the yoke of the magnetic write-head distributes among these two channels, they should be represented by two parallel reluctances, $R_g$ and $R_s$, as shown in Fig. \ref{ZL_Model} (b). The gap reluctance, $R_g$, follows the conventional geometrical definition of reluctance
\begin{equation}\label{Rg}
R_g=\frac{\l_g}{\mu_0 A_{gap}}=\frac{\l_g}{\mu_0 l_y d_{yoke}}  \qquad  H^{-1}
\end{equation}
where $A_{gap}$ is the effective cross-sectional area of stray flux in the air gap, $d_{yoke}$ is the effective thickness of the stray flux channel inside the air gap ( Note that $d_{yoke}$ is analogous to the thickness of the magnetic yoke in the ring shape head for longitudinal recording ), $l_y$ is the the size of the yoke in the transverse dimensions (see Fig. \ref{ZL_Model}). The Meissner state reluctance $R_s$ is generated due to the penetration of magnetic field into the superconductor within its Meissner screening region. Therefore, the total flux will shunt into two branches, $\Phi_s$ with its reluctance $R_s$ and $\Phi_g$ with its reluctance $R_g$. In the linear Meissner state, we obtain

\begin{equation}
\begin{split}
\Phi_g= B * A_{gap} = B*(l_y * d_{yoke}) \\
\Phi_s = B*(l_y * \lambda(T))
\end{split}
\end{equation}

where $B$ is the amplitude of the applied RF magnetic field in the gap (assumed to be the same field witnessed by the sample) and $\lambda(T)$ is the temperature dependent penetration depth. By applying the node-voltage law, we have $\Phi_g$ $R_g$ = $\Phi_s$ $R_s$. Hence, the reluctance of the linear Meissner state is given by
\begin{equation}\label{Rs}
R_s=\frac{d_{yoke}}{\lambda(T)} R_g  \qquad where \; \; R_g=\frac{\l_g}{\mu_0 l_y d_{yoke}}
\end{equation}

So the total reluctance in this magnetic circuit is
\begin{equation}
\begin{split}
R_{eq}=R_y+R_g//R_s =R_y+ \frac{d_{yoke}}{d_{yoke}+\lambda}R_g \\
=\left(\frac{\mu_0}{\mu_f}+\frac{d_{yoke}}{d_{yoke}+\lambda}\right) R_g
\end{split}
\end{equation}
where $R_y$ is the reluctance of the magnetic yoke and can be approximated as $\frac{\mu_0}{\mu_f}R_g$, here $\mu_0$ and $\mu_f$ indicate the permeability of the vacuum and the magnetic yoke, respectively. By solving the magnetic circuit in Fig. \ref{ZL_Model}, one finds the magnetic flux
flowing through the magnetic circuit will, in turn, give the induced electric voltage into
the input winding through Faraday's law
\begin{equation}
\begin{split}
V_{e}=\frac{d \Phi}{dt}=\frac{d}{dt}\left(\frac{V_m}{R_{eq}}\right) =\frac{d}{dt} (\frac{NI}{R_{eq}}) \\
= \frac{iN \omega I_0 e^{i \omega t}/R_g}{\mu_0/\mu_f + d_{yoke}/(d_{yoke}+\lambda(T))}
\end{split}
\end{equation}
where $V_m$ is the impressed voltage on the magnetic circuit and can be written as $V_m=NI$. Here N is the number of turns in the solenoid and $I$ indicates the RF input current with a maximum amplitude $I_0$ in an RF cycle ($I=I_0 e^{i \omega t}$). Therefore the impedance of the magnetic circuit can be found as
$\frac{iN \omega /R_g}{\mu_0/\mu_f + d_{yoke}/(d_{yoke}+\lambda(T))}$.
To evaluate the microwave load impedance presented by the write head, $Z_L$ , one should include the impedance coming from the rest of the probe circuit to get
\begin{equation}\label{ZL}
Z_L=Z_{Bkd}+\frac{i \omega N/ R_g }{\frac{\mu_0}{\mu_f} +\frac{d_{yoke}}{d_{yoke}+\lambda(T)}}  \; ; \; Z_{Bkd} \equiv R_{Bkd}+i X_{Bkd}
\end{equation}
where $R_{Bkd}$ and $X_{Bkd}$ represent the background resistance and reactance of the load impedance, respectively (i.e. these parameters model the background probe-dependent properties of the write head). Note that the load impedance model does not include the loss in the superconductor. In addition, while the film thickness is smaller than the penetration depth, the regular penetration depth should be replaced by an effective penetration depth, $\lambda_{e}$ \cite{TinkhamBook}
\begin{equation} \label{effictive_lambda}
\lambda(T)\rightarrow \lambda_{e}(T)=\frac{\lambda^2(T)} {d_{film}}
\end{equation}
where $d_{film}$ is the thickness of the film and it is assumed that $d_{film} \ll \lambda$.

\begin{figure}
\center
\includegraphics*[width=2.0in]{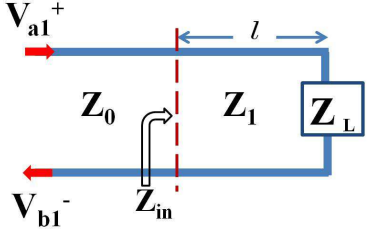}
\caption{Model the electrical impedance looking into a magnetic write head probe at the end of a short segment of transmission line. $Z_0$ indicates the characteristic impedance of the calibrated transmission line (50 $\Omega$). $Z_1$ is the characteristic impedance of the probe transmission line (usually not 50 $\Omega$) with length $l$. $Z_L$ represents the load impedance including the impedance of the magnetic probe head and a superconducting sample. $V_{a1}^{+}$ and $V_{b1}^{-}$ indicate the input and reflected microwave tones respectively.}
\label{S11_Model}
\end{figure}

In the experiment schematically shown in Fig. \ref{S11Setup}, the probe load impedance (represented now by $Z_{L}$) is connected to a printed circuit board and then a semi-rigid coax cable of finite length is soldered to the contacts of the printed circuit board. The length of the printed circuit board plus the length of semi-rigid coax cable can be combined together and treated as a transmission line of length $l$ on the probe assembly. One can model the microwave circuit of a transmission line terminated by $Z_L$ as shown in Fig. \ref{S11_Model}. In this one port measurement, the microwave calibration is done only on the transmission line from the part of the flexible coax cable shown in Fig. \ref{S11Setup} but not on the probe assembly. A characteristic impedance mismatch between the calibrated transmission line and the un-calibrated part will be inevitable. Here $Z_0$ represents the characteristic impedance of the calibrated transmission line, 50 Ohms in our case, and $Z_1$ is the characteristic impedance of the un-calibrated transmission line on the probe assembly with a length of $l$. Both $Z_0$ and $Z_1$ are real.
In this circumstance, the input impedance, $Z_{in}$ is the impedance seen at the plane of calibration, given by,
\begin{equation}
Z_{in}=Z_1\frac{Z_L+ j Z_1 tan(\gamma l)}{Z_1 + j Z_L  tan(\gamma l) } \;\; where \;\; \gamma=\beta -  i \alpha= \frac{n \omega}{c} - i \alpha
\end{equation}
Here $n$ is the refractive index of the transmission line of length $l$, $\omega$ is the angular frequency and $\alpha$ is an attenuation coefficient inside the transmission line. Finally the theoretical $S_{11}$ based on this model can be calculated as
\begin{equation}
S_{11}=\frac{Z_{in}-Z_0}{Z_{in}+Z_0},
\end{equation}
and compared to experiment.

Complex curve fitting of this model to the $S_{11}(T)$ data is performed by the least square method to minimize the difference between the experimental temperature dependent $S_{11}$ values and the values predicted by the model. The details of the complex curve fitting method for the linear response measurement are discussed in Ref. \cite{Tai_thesis}.

For the curve fitting, the index of refraction of the Teflon inside the coaxial cable in our experiment is taken to be $n$ $\approx$ 1.2, independent of the excitation frequency. The un-calibrated length of the transmission line is around 5$\sim$6 cm. Conveniently, we combine the $n$ and $l$ together to treat them as a constant term $nl=0.065$ $m$ in the calculation. On the other hand, the coefficient $\alpha$ (unit: $m^{-1}$) and $l$ (unit: m) are bundled together as a dimensionless fitting parameter. The effective thickness of stray field in the air gap is assumed to be a constant, $d_{yoke}$=500 nm. The temperature dependence of an effective penetration depth, $\lambda_e$, is approximately described by \cite{TinkhamBook}
\begin{equation}
\lambda_e(T)\approx\frac{\left( \lambda(0)\Big[1-(T/T_c)^4\Big]^{-\frac{1}{2}} \right)^2}{d_{film}}
\end{equation}
where $\lambda(0)=40$ $nm$ is the $0$ $K$ penetration depth of Nb and $d_{film}=50$ $nm$ is the known thickness of the measured Nb thin film. The transition temperature $T_c$=8.3 K is kept fixed for the fit. Therefore, based on the model, the unknown coefficients in the experiment are $\alpha l$, $Z_1$, $R_{Bkd}$, $X_{Bkd}$, $N/R_g$ and $\mu_0 / \mu_f$.

The fit is done on the data set measured by the PINNACLE probe on the 50 nm Nb film with an excitation frequency of 1.98 GHz as discussed above. The solid lines in Fig. \ref{NbS11T} (a) and Fig.\ref{NbS11T} (b) show the theoretical amplitude and phase respectively, based on the model with the following fit coefficients : $\alpha l=0.245$ (unit-less), $Z_1=30.23$ $\Omega$, $R_{Bkd}=320.48$ $\Omega$, $X_{Bkd}=-105.13$ $\Omega$, $N/R_g=6.59*10^{-9}$ $Henry$ and $\mu_0 / \mu_f =-0.46$. Again, the experimental data for the amplitude and phase is indicated in the figure by dot circles and the fit is excellent. Note that the $\mu_0 /\mu_f$ is a negative value, which implies that the excitation frequency is above and not far away from the yoke ferrite ferromagnetic resonance frequency. All of the fitting coefficients are physically reasonable. The sensitivity of the fits to variation of the fixed parameters in the fitting was also studied. For example, the variation of $d_{yoke}$ from 200 nm to 1 $\mu m$ will make a change in the $\mu_0 /\mu_f$, $R_{Bkd}$, and $X_{Bkd}$ values with $\pm 15 \% $ variation.  The $N/R_g$ parameter varies by $\pm 1 \% $ and is on the order of $10^{-9}$ $Henry$, not very sensitive to the change of $d_{yoke}$ and the rest of the fixed parameters. Finally, note that the model is designed only to fit the data in the superconducting state. In the normal state, the penetration depth $\lambda(T)$ in Eq. \ref{ZL} should change to the skin depth of the normal state of Nb. In this case, the field configuration around the sample is very different, and the situation is of not of interest to us here.

\section{Discussion}

With this model, the magnitude and temperature dependence of the magnetic penetration depth of any superconducting thin film can be calculated after all of the parameters in the fit are calibrated by a standard thin film.  Then a raster scan on the surface of superconductor at constant height will successfully image the contrast of $\lambda$ of any superconducting thin film. The lateral length scales measured with this probe are on the order of the gap length $l_g$, or probe-sample separation, whichever is greater \cite{Tai2012}\cite{Tai2013}. However compared with a nonlinearity measurement, the linear response measurement is not as sensitive to the extrinsic electromagnetic response associated with defects on superconductors \cite{Tai2013}\cite{Tai2014}\cite{Oates2007}. Another limitation for the linear response measurement is the fact that we could not develop signals from bulk Nb surfaces. The reason is that the
surface reactance in thin films is more significant
than that of bulk superconductors, as illustrated in Eq. \ref{effictive_lambda}, which leads to a
measurable change in the linear response of Nb thin films but not on bulk Nb. This limitation further motivates the need to pursue localized nonlinear measurements to get information about bulk superconductors \cite{Tai2011}\cite{Tai2012}\cite{Tai2013}\cite{Tai2014}.\\

\section{Conclusions}

From the linear response measurement of Nb thin films
by a magnetic write head probe, the electrodynamic properties of Nb can be identified.
The linear response can also be used to find the film $T_c$ in a local area. A magnetic circuit model combined with the transmission line circuit is given to interpret the linear response measurement. From the curve fitting results, many unknown coefficients in our microwave circuit can be fixed. This model will enable quantitatively imaging of the microwave surface properties of superconducting materials.

\section{Acknowledgements}
This work is supported by the US Department of Energy $/$ High Energy
Physics through grant $\#$ DESC0004950, and also by the ONR AppEl,
Task D10, (Award No.\ N000140911190), and CNAM.

\bibliography{Tai_JAPbibliography}

\begin{thebibliography}{29}
\expandafter\ifx\csname natexlab\endcsname\relax\def\natexlab#1{#1}\fi
\expandafter\ifx\csname bibnamefont\endcsname\relax
  \def\bibnamefont#1{#1}\fi
\expandafter\ifx\csname bibfnamefont\endcsname\relax
  \def\bibfnamefont#1{#1}\fi
\expandafter\ifx\csname citenamefont\endcsname\relax
  \def\citenamefont#1{#1}\fi
\expandafter\ifx\csname url\endcsname\relax
  \def\url#1{\texttt{#1}}\fi
\expandafter\ifx\csname urlprefix\endcsname\relax\def\urlprefix{URL }\fi
\providecommand{\bibinfo}[2]{#2}
\providecommand{\eprint}[2][]{\url{#2}}

\bibitem[{\citenamefont{Talanov et~al.}(2000)\citenamefont{Talanov, Mercaldo,
  Anlage, and Claassen}}]{Talanov}
\bibinfo{author}{\bibfnamefont{V.~V.} \bibnamefont{Talanov}},
  \bibinfo{author}{\bibfnamefont{L.~V.} \bibnamefont{Mercaldo}},
  \bibinfo{author}{\bibfnamefont{S.~M.} \bibnamefont{Anlage}},
  \bibnamefont{and} \bibinfo{author}{\bibfnamefont{J.~H.}
  \bibnamefont{Claassen}}, \bibinfo{journal}{Review of Scientific Instruments}
  \textbf{\bibinfo{volume}{71}}, \bibinfo{pages}{2136} (\bibinfo{year}{2000}).

\bibitem[{\citenamefont{Tinkham}(2004)}]{TinkhamBook}
\bibinfo{author}{\bibfnamefont{M.}~\bibnamefont{Tinkham}},
  \emph{\bibinfo{title}{Introduction to Superconductivity}}
  (\bibinfo{publisher}{McGraw-Hill}, \bibinfo{address}{New York},
  \bibinfo{year}{2004}), \bibinfo{edition}{2nd} ed.

\bibitem[{\citenamefont{Barone and Patern$\grave{o}$}(1982)}]{Barone}
\bibinfo{author}{\bibfnamefont{A.}~\bibnamefont{Barone}} \bibnamefont{and}
  \bibinfo{author}{\bibfnamefont{G.}~\bibnamefont{Patern$\grave{o}$}},
  \emph{\bibinfo{title}{Physics and Applications of Josephson Effect}}
  (\bibinfo{publisher}{Wiley}, \bibinfo{year}{1982}).

\bibitem[{\citenamefont{Cunnane et~al.}(2013)\citenamefont{Cunnane, Zhuang,
  Chen, Xi, Yong, and Lemberger}}]{Cunnane2013}
\bibinfo{author}{\bibfnamefont{D.}~\bibnamefont{Cunnane}},
  \bibinfo{author}{\bibfnamefont{C.}~\bibnamefont{Zhuang}},
  \bibinfo{author}{\bibfnamefont{K.}~\bibnamefont{Chen}},
  \bibinfo{author}{\bibfnamefont{X.~X.} \bibnamefont{Xi}},
  \bibinfo{author}{\bibfnamefont{J.}~\bibnamefont{Yong}}, \bibnamefont{and}
  \bibinfo{author}{\bibfnamefont{T.~R.} \bibnamefont{Lemberger}},
  \bibinfo{journal}{Appl. Phys. Lett.} \textbf{\bibinfo{volume}{102}},
  \bibinfo{pages}{072603} (\bibinfo{year}{2013}).

\bibitem[{\citenamefont{Klein et~al.}(1989)\citenamefont{Klein, Muller, Piel,
  Roas, Shultz, Klein, and Peininger}}]{Klein1989}
\bibinfo{author}{\bibfnamefont{N.}~\bibnamefont{Klein}},
  \bibinfo{author}{\bibfnamefont{G.}~\bibnamefont{Muller}},
  \bibinfo{author}{\bibfnamefont{H.}~\bibnamefont{Piel}},
  \bibinfo{author}{\bibfnamefont{B.}~\bibnamefont{Roas}},
  \bibinfo{author}{\bibfnamefont{L.}~\bibnamefont{Shultz}},
  \bibinfo{author}{\bibfnamefont{U.}~\bibnamefont{Klein}}, \bibnamefont{and}
  \bibinfo{author}{\bibfnamefont{M.}~\bibnamefont{Peininger}},
  \bibinfo{journal}{Appl. Phys. Lett.} \textbf{\bibinfo{volume}{54}},
  \bibinfo{pages}{757} (\bibinfo{year}{1989}).

\bibitem[{\citenamefont{Klein}(1992)}]{Klein1992}
\bibinfo{author}{\bibfnamefont{N.}~\bibnamefont{Klein}}, \bibinfo{journal}{J.
  Supercond.} \textbf{\bibinfo{volume}{5}}, \bibinfo{pages}{195}
  (\bibinfo{year}{1992}).

\bibitem[{\citenamefont{Wilker et~al.}(1992)\citenamefont{Wilker, Shen, Nguen,
  and Brenner}}]{Wilker1992}
\bibinfo{author}{\bibfnamefont{C.}~\bibnamefont{Wilker}},
  \bibinfo{author}{\bibfnamefont{Z.-Y.} \bibnamefont{Shen}},
  \bibinfo{author}{\bibfnamefont{V.-X.} \bibnamefont{Nguen}}, \bibnamefont{and}
  \bibinfo{author}{\bibfnamefont{M.~S.} \bibnamefont{Brenner}},
  \bibinfo{journal}{IEEE Trans. Appl. Supercond.} \textbf{\bibinfo{volume}{3}},
  \bibinfo{pages}{2832} (\bibinfo{year}{1992}).

\bibitem[{\citenamefont{Kobayashi and Yoshikawa}(1998)}]{Kobayashi}
\bibinfo{author}{\bibfnamefont{Y.}~\bibnamefont{Kobayashi}} \bibnamefont{and}
  \bibinfo{author}{\bibfnamefont{H.}~\bibnamefont{Yoshikawa}},
  \bibinfo{journal}{Microwave Theory Tech.} \textbf{\bibinfo{volume}{46}},
  \bibinfo{pages}{2524} (\bibinfo{year}{1998}).

\bibitem[{\citenamefont{Luan et~al.}(2010)\citenamefont{Luan, Auslaender,
  Lippman, Hicks, Kalisky, Chu, Analytis, Fisher, Kirtley, and
  Moler}}]{Moler2010}
\bibinfo{author}{\bibfnamefont{L.}~\bibnamefont{Luan}},
  \bibinfo{author}{\bibfnamefont{O.~M.} \bibnamefont{Auslaender}},
  \bibinfo{author}{\bibfnamefont{T.~M.} \bibnamefont{Lippman}},
  \bibinfo{author}{\bibfnamefont{C.~W.} \bibnamefont{Hicks}},
  \bibinfo{author}{\bibfnamefont{B.}~\bibnamefont{Kalisky}},
  \bibinfo{author}{\bibfnamefont{J.-H.} \bibnamefont{Chu}},
  \bibinfo{author}{\bibfnamefont{J.~G.} \bibnamefont{Analytis}},
  \bibinfo{author}{\bibfnamefont{I.~R.} \bibnamefont{Fisher}},
  \bibinfo{author}{\bibfnamefont{J.~R.} \bibnamefont{Kirtley}},
  \bibnamefont{and} \bibinfo{author}{\bibfnamefont{K.~A.} \bibnamefont{Moler}},
  \bibinfo{journal}{Phys. Rev. B} \textbf{\bibinfo{volume}{81}},
  \bibinfo{pages}{100501} (\bibinfo{year}{2010}).

\bibitem[{\citenamefont{Hicks et~al.}(2009)\citenamefont{Hicks, Lippman, Huber,
  Analytis, Chu, Erickson, Fisher, and Moler}}]{Moler2009}
\bibinfo{author}{\bibfnamefont{C.~W.} \bibnamefont{Hicks}},
  \bibinfo{author}{\bibfnamefont{T.~M.} \bibnamefont{Lippman}},
  \bibinfo{author}{\bibfnamefont{M.~E.} \bibnamefont{Huber}},
  \bibinfo{author}{\bibfnamefont{J.~G.} \bibnamefont{Analytis}},
  \bibinfo{author}{\bibfnamefont{J.-H.} \bibnamefont{Chu}},
  \bibinfo{author}{\bibfnamefont{A.~S.} \bibnamefont{Erickson}},
  \bibinfo{author}{\bibfnamefont{I.~R.} \bibnamefont{Fisher}},
  \bibnamefont{and} \bibinfo{author}{\bibfnamefont{K.~A.} \bibnamefont{Moler}},
  \bibinfo{journal}{Phys. Rev. Lett.} \textbf{\bibinfo{volume}{103}},
  \bibinfo{pages}{127003} (\bibinfo{year}{2009}).

\bibitem[{\citenamefont{Kirtley et~al.}(2012)\citenamefont{Kirtley, Kalisky,
  Bert, Bell, Kim, Hikita, Hwang, Ngai, Segal, Walker et~al.}}]{Moler2012}
\bibinfo{author}{\bibfnamefont{J.~R.} \bibnamefont{Kirtley}},
  \bibinfo{author}{\bibfnamefont{B.}~\bibnamefont{Kalisky}},
  \bibinfo{author}{\bibfnamefont{J.~A.} \bibnamefont{Bert}},
  \bibinfo{author}{\bibfnamefont{C.}~\bibnamefont{Bell}},
  \bibinfo{author}{\bibfnamefont{M.}~\bibnamefont{Kim}},
  \bibinfo{author}{\bibfnamefont{Y.}~\bibnamefont{Hikita}},
  \bibinfo{author}{\bibfnamefont{H.~Y.} \bibnamefont{Hwang}},
  \bibinfo{author}{\bibfnamefont{J.~H.} \bibnamefont{Ngai}},
  \bibinfo{author}{\bibfnamefont{Y.}~\bibnamefont{Segal}},
  \bibinfo{author}{\bibfnamefont{F.~J.} \bibnamefont{Walker}},
  \bibnamefont{et~al.}, \bibinfo{journal}{Phys. Rev. B}
  \textbf{\bibinfo{volume}{85}}, \bibinfo{pages}{224518}
  (\bibinfo{year}{2012}).

\bibitem[{\citenamefont{Padamsee et~al.}(1998)\citenamefont{Padamsee, Knobloch,
  and Hays}}]{Padamsee}
\bibinfo{author}{\bibfnamefont{H.}~\bibnamefont{Padamsee}},
  \bibinfo{author}{\bibfnamefont{J.}~\bibnamefont{Knobloch}}, \bibnamefont{and}
  \bibinfo{author}{\bibfnamefont{T.}~\bibnamefont{Hays}},
  \emph{\bibinfo{title}{RF Superconductivities on Accelerators}}
  (\bibinfo{publisher}{Wiley Series in Beam Physics and Accelerator
  Technology}, \bibinfo{year}{1998}).

\bibitem[{\citenamefont{Vlahacos et~al.}(2011)\citenamefont{Vlahacos, Matthews,
  and Wellstood}}]{Wellstood}
\bibinfo{author}{\bibfnamefont{C.~P.} \bibnamefont{Vlahacos}},
  \bibinfo{author}{\bibfnamefont{J.}~\bibnamefont{Matthews}}, \bibnamefont{and}
  \bibinfo{author}{\bibfnamefont{F.~C.} \bibnamefont{Wellstood}},
  \bibinfo{journal}{IEEE Trans. Appl. Supercond.}
  \textbf{\bibinfo{volume}{21}}, \bibinfo{pages}{412} (\bibinfo{year}{2011}).

\bibitem[{\citenamefont{Ciovati et~al.}(2010)\citenamefont{Ciovati, Myneni,
  Stevie, Maheshwari, and Griffis}}]{Ciovati2010}
\bibinfo{author}{\bibfnamefont{G.}~\bibnamefont{Ciovati}},
  \bibinfo{author}{\bibfnamefont{G.}~\bibnamefont{Myneni}},
  \bibinfo{author}{\bibfnamefont{F.}~\bibnamefont{Stevie}},
  \bibinfo{author}{\bibfnamefont{P.}~\bibnamefont{Maheshwari}},
  \bibnamefont{and} \bibinfo{author}{\bibfnamefont{D.}~\bibnamefont{Griffis}},
  \bibinfo{journal}{Phys. Rev. Special Topics-Accerlerators and Beams}
  \textbf{\bibinfo{volume}{13}}, \bibinfo{pages}{022002}
  (\bibinfo{year}{2010}).

\bibitem[{\citenamefont{Xi}()}]{Xi2009}
\bibinfo{author}{\bibfnamefont{X.~X.} \bibnamefont{Xi}},
  \bibinfo{journal}{Supercond. Sci. Technol.} \textbf{\bibinfo{volume}{22}},
  \bibinfo{pages}{043001} (????).

\bibitem[{\citenamefont{Tajima et~al.}(2007)\citenamefont{Tajima, Canabal,
  Zhao, Romanenko, Moeckly, Nantista, Tantawi, Phillips, Iwashita, and
  Campisi}}]{Tajima}
\bibinfo{author}{\bibfnamefont{T.}~\bibnamefont{Tajima}},
  \bibinfo{author}{\bibfnamefont{A.}~\bibnamefont{Canabal}},
  \bibinfo{author}{\bibfnamefont{Y.}~\bibnamefont{Zhao}},
  \bibinfo{author}{\bibfnamefont{A.}~\bibnamefont{Romanenko}},
  \bibinfo{author}{\bibfnamefont{B.~H.} \bibnamefont{Moeckly}},
  \bibinfo{author}{\bibfnamefont{C.~D.} \bibnamefont{Nantista}},
  \bibinfo{author}{\bibfnamefont{S.}~\bibnamefont{Tantawi}},
  \bibinfo{author}{\bibfnamefont{L.}~\bibnamefont{Phillips}},
  \bibinfo{author}{\bibfnamefont{Y.}~\bibnamefont{Iwashita}}, \bibnamefont{and}
  \bibinfo{author}{\bibfnamefont{I.~E.} \bibnamefont{Campisi}},
  \bibinfo{journal}{IEEE Trans. Appl. Supercond.}
  \textbf{\bibinfo{volume}{17}}, \bibinfo{pages}{1330} (\bibinfo{year}{2007}).

\bibitem[{\citenamefont{Tai et~al.}(2011)\citenamefont{Tai, Xi, Zhuang, Mircea,
  and Anlage}}]{Tai2011}
\bibinfo{author}{\bibfnamefont{T.}~\bibnamefont{Tai}},
  \bibinfo{author}{\bibfnamefont{X.~X.} \bibnamefont{Xi}},
  \bibinfo{author}{\bibfnamefont{C.~G.} \bibnamefont{Zhuang}},
  \bibinfo{author}{\bibfnamefont{D.~I.} \bibnamefont{Mircea}},
  \bibnamefont{and} \bibinfo{author}{\bibfnamefont{S.~M.}
  \bibnamefont{Anlage}}, \bibinfo{journal}{IEEE Trans. Appl. Supercond.}
  \textbf{\bibinfo{volume}{21}}, \bibinfo{pages}{2615} (\bibinfo{year}{2011}).

\bibitem[{\citenamefont{Tai et~al.}(2012)\citenamefont{Tai, Ghamsari, Tan,
  Zhuang, Xi, and Anlage}}]{Tai2012}
\bibinfo{author}{\bibfnamefont{T.}~\bibnamefont{Tai}},
  \bibinfo{author}{\bibfnamefont{B.~G.} \bibnamefont{Ghamsari}},
  \bibinfo{author}{\bibfnamefont{T.}~\bibnamefont{Tan}},
  \bibinfo{author}{\bibfnamefont{C.~G.} \bibnamefont{Zhuang}},
  \bibinfo{author}{\bibfnamefont{X.~X.} \bibnamefont{Xi}}, \bibnamefont{and}
  \bibinfo{author}{\bibfnamefont{S.~M.} \bibnamefont{Anlage}},
  \bibinfo{journal}{Phys. Rev. ST Accel. Beams} \textbf{\bibinfo{volume}{15}},
  \bibinfo{pages}{122002} (\bibinfo{year}{2012}).

\bibitem[{\citenamefont{Tai et~al.}(2013{\natexlab{a}})\citenamefont{Tai,
  Ghamsari, and Anlage}}]{Tai2013}
\bibinfo{author}{\bibfnamefont{T.}~\bibnamefont{Tai}},
  \bibinfo{author}{\bibfnamefont{B.~G.} \bibnamefont{Ghamsari}},
  \bibnamefont{and} \bibinfo{author}{\bibfnamefont{S.~M.}
  \bibnamefont{Anlage}}, \bibinfo{journal}{IEEE Trans. Appl. Supercond.}
  \textbf{\bibinfo{volume}{23}}, \bibinfo{pages}{7100104}
  (\bibinfo{year}{2013}{\natexlab{a}}).

\bibitem[{\citenamefont{Tai et~al.}(2013{\natexlab{b}})\citenamefont{Tai,
  Ghamsari, Bieler, Tan, Xi, and Anlage}}]{Tai2014}
\bibinfo{author}{\bibfnamefont{T.}~\bibnamefont{Tai}},
  \bibinfo{author}{\bibfnamefont{B.~G.} \bibnamefont{Ghamsari}},
  \bibinfo{author}{\bibfnamefont{T.~R.} \bibnamefont{Bieler}},
  \bibinfo{author}{\bibfnamefont{T.}~\bibnamefont{Tan}},
  \bibinfo{author}{\bibfnamefont{X.~X.} \bibnamefont{Xi}}, \bibnamefont{and}
  \bibinfo{author}{\bibfnamefont{S.~M.} \bibnamefont{Anlage}}
  (\bibinfo{year}{2013}{\natexlab{b}}), \eprint{{arXiv:1312.6257v1}}.

\bibitem[{\citenamefont{O'Barr et~al.}(1996)\citenamefont{O'Barr, Lederman, and
  Schultz}}]{Schultz1996}
\bibinfo{author}{\bibfnamefont{R.}~\bibnamefont{O'Barr}},
  \bibinfo{author}{\bibfnamefont{M.}~\bibnamefont{Lederman}}, \bibnamefont{and}
  \bibinfo{author}{\bibfnamefont{S.}~\bibnamefont{Schultz}},
  \bibinfo{journal}{J. Appl. Phys.} \textbf{\bibinfo{volume}{79}},
  \bibinfo{pages}{6069} (\bibinfo{year}{1996}).

\bibitem[{\citenamefont{Yamamoto and Schultz}(1996)}]{Yamamoto1996}
\bibinfo{author}{\bibfnamefont{S.~Y.} \bibnamefont{Yamamoto}} \bibnamefont{and}
  \bibinfo{author}{\bibfnamefont{S.}~\bibnamefont{Schultz}},
  \bibinfo{journal}{Appl. Phys. Lett.} \textbf{\bibinfo{volume}{69}},
  \bibinfo{pages}{3263} (\bibinfo{year}{1996}).

\bibitem[{\citenamefont{Yamamoto and Schultz}(1997)}]{Yamamoto1997}
\bibinfo{author}{\bibfnamefont{S.~Y.} \bibnamefont{Yamamoto}} \bibnamefont{and}
  \bibinfo{author}{\bibfnamefont{S.}~\bibnamefont{Schultz}},
  \bibinfo{journal}{J. Appl. Phys.} \textbf{\bibinfo{volume}{81}},
  \bibinfo{pages}{4696} (\bibinfo{year}{1997}).

\bibitem[{\citenamefont{Pond et~al.}(1997)\citenamefont{Pond, Allen, and
  Cukauskas}}]{Pond}
\bibinfo{author}{\bibfnamefont{J.~M.} \bibnamefont{Pond}},
  \bibinfo{author}{\bibfnamefont{L.~H.} \bibnamefont{Allen}}, \bibnamefont{and}
  \bibinfo{author}{\bibfnamefont{E.~J.} \bibnamefont{Cukauskas}},
  \bibinfo{journal}{IEEE Trans. Appl. Supercond.} \textbf{\bibinfo{volume}{7}},
  \bibinfo{pages}{1857} (\bibinfo{year}{1997}).

\bibitem[{\citenamefont{Liu et~al.}(2001)\citenamefont{Liu, Shi, Wang, Chen,
  Stoev, Leal, Saha, Tong, Dey, and Nojaba}}]{ReadRiteCorp}
\bibinfo{author}{\bibfnamefont{F.~H.} \bibnamefont{Liu}},
  \bibinfo{author}{\bibfnamefont{S.}~\bibnamefont{Shi}},
  \bibinfo{author}{\bibfnamefont{J.}~\bibnamefont{Wang}},
  \bibinfo{author}{\bibfnamefont{Y.}~\bibnamefont{Chen}},
  \bibinfo{author}{\bibfnamefont{K.}~\bibnamefont{Stoev}},
  \bibinfo{author}{\bibfnamefont{L.}~\bibnamefont{Leal}},
  \bibinfo{author}{\bibfnamefont{R.}~\bibnamefont{Saha}},
  \bibinfo{author}{\bibfnamefont{H.}~\bibnamefont{Tong}},
  \bibinfo{author}{\bibfnamefont{S.}~\bibnamefont{Dey}}, \bibnamefont{and}
  \bibinfo{author}{\bibfnamefont{M.}~\bibnamefont{Nojaba}},
  \bibinfo{journal}{IEEE Trans. Appl.} \textbf{\bibinfo{volume}{37}},
  \bibinfo{pages}{613} (\bibinfo{year}{2001}).

\bibitem[{\citenamefont{Koblischka et~al.}(2007)\citenamefont{Koblischka, Wei,
  Sulzbach, Johnston, and Hartmann}}]{Koblischka2007}
\bibinfo{author}{\bibfnamefont{M.~R.} \bibnamefont{Koblischka}},
  \bibinfo{author}{\bibfnamefont{J.-D.} \bibnamefont{Wei}},
  \bibinfo{author}{\bibfnamefont{T.}~\bibnamefont{Sulzbach}},
  \bibinfo{author}{\bibfnamefont{A.~D.} \bibnamefont{Johnston}},
  \bibnamefont{and} \bibinfo{author}{\bibfnamefont{U.}~\bibnamefont{Hartmann}},
  \bibinfo{journal}{IEEE Trans. on Magnetics} \textbf{\bibinfo{volume}{43}},
  \bibinfo{pages}{2205} (\bibinfo{year}{2007}).

\bibitem[{\citenamefont{Koblischka et~al.}(2010)\citenamefont{Koblischka, Wei,
  and Hartmann}}]{Koblischka2010}
\bibinfo{author}{\bibfnamefont{M.~R.} \bibnamefont{Koblischka}},
  \bibinfo{author}{\bibfnamefont{J.~D.} \bibnamefont{Wei}}, \bibnamefont{and}
  \bibinfo{author}{\bibfnamefont{U.}~\bibnamefont{Hartmann}},
  \bibinfo{journal}{J. Magn. Magn. Mater.} \textbf{\bibinfo{volume}{322}},
  \bibinfo{pages}{1694} (\bibinfo{year}{2010}).

\bibitem[{\citenamefont{Tai}(2013)}]{Tai_thesis}
\bibinfo{author}{\bibfnamefont{T.}~\bibnamefont{Tai}},
  \emph{\bibinfo{title}{Measuring electromagnetic properties of superconductors
  in high and localized RF magnetic field}} (\bibinfo{publisher}{Ph.D
  dissertation; Univ. Maryland-College Park},
  \bibinfo{address}{http://hdl.handle.net/1903/14668}, \bibinfo{year}{2013}).

\bibitem[{\citenamefont{Oates et~al.}(2007)\citenamefont{Oates, Agassi, and
  Moeckly}}]{Oates2007}
\bibinfo{author}{\bibfnamefont{D.~E.} \bibnamefont{Oates}},
  \bibinfo{author}{\bibfnamefont{Y.~D.} \bibnamefont{Agassi}},
  \bibnamefont{and} \bibinfo{author}{\bibfnamefont{B.~H.}
  \bibnamefont{Moeckly}}, \bibinfo{journal}{IEEE Trans. Appl. Supercond.}
  \textbf{\bibinfo{volume}{17}}, \bibinfo{pages}{2871} (\bibinfo{year}{2007}).

\end{thebibliography}
\end{document}